# Spin-current-related magnetoresistance in epitaxial Pt/Co bilayers in the presence of spin Hall effect and Rashba-Edelstein effect

Ye Du[1,2], Saburo Takahashi[3], and Junsaku Nitta[1,2,4,*]

[1]Department of Materials Science, Tohoku University, Sendai 980-8579, Japan

[2]Center for Science and Innovation in Spintronics, Tohoku University, Sendai 980-8577, Japan

[3]Advanced Institute for Materials Research, Tohoku University, Sendai 980-8577

[4]Center for Spintronics Research Network, Tohoku University, Sendai 980-8577, Japan

## Abstract

We analyze the experimentally obtained spin-current-related magnetoresistance in epitaxial Pt/Co bilayers by using a drift-diffusion model that incorporates both bulk spin Hall effect and interfacial Rashba-Edelstein effect (REE). The magnetoresistance analysis yields, for the Pt/Co interface, a temperature-independent Rashba parameter in the order of $10^{-11}$ eV m that agrees with theoretical calculations, along with an effective interfacial REE thickness of several angstroms which is in overall consistency with our previous spin-orbit torque analysis. In particular, our results suggest that both bulk and interface charge-spin current inter-conversions need to be taken into account for the spin-current-related magnetoresistance analysis in highly-conductive magnetic hetero-structures such as the epitaxial Pt/Co bilayers.

# I. INTRODUCTION

The giant spin Hall effect (SHE) [1] has been shown in recent years to couple to magnetization dynamics, giving rise to rich physical phenomena such as the ferromagnetic resonance-based spin pumping [2], the magnetic damping modulation [3], the spin-orbit spin-transfer torque [4] and the spin-orbit magnetization switching [5–7]. Recently, the SHE was also shown to cause a resistance change in heavy-metal/ferromagnet bilayers depending on the magnetization direction, namely, the spin Hall magnetoresistance (SMR) [8–11], which originates from a joint action of the direct and inverse SHE. The SMR is shown to be an effective method for probing the spin-related properties in the heavy-metal layer such as the spin Hall angle and the spin diffusion length [9,11,12].

In contrast to SMR which stems from the bulk SHE, there is another type of spin-related MR, the Rashba-Edelstein magnetoresistance (REMR) [13], which originates from the Rashba-Edelstein effect (REE) [14–16] that occurs at hetero-interface with broken inversion symmetry. Recently, the REMR was experimentally reported in Bi/Ag/ferromagnet multilayers [13,17], arising from a non-equilibrium spin accumulation at Bi/Ag interface [18] caused by the giant spin splitting [19]. With spin current diffusing across the Ag/ferromagnet interface, the resistance of the Bi/Ag/ferromagnet multilayers changes depending on transmission or reflection of the spin current near Ag/ferromagnet interface, resembling the mechanism of SMR [8,9].

In metallic heavy-metal/ferromagnet bilayers and multilayers, due to the strong spin-orbit coupling in the heavy-metal layer, the SHE is often considered as the source of spin-current generation for the SMR analysis, for instance in the case of Pt-based [20–23], Ta-based [11,24,25] and W-based [10,11,26] hetero-structures. However, theoretical studies have also shown that giant spin splitting is present at heavy-metal (*e.g.* Pt)/ferromagnet interfaces [27,28], giving rise to a large Rashba parameter in the order of $10^{-11}$–$10^{-10}$ eV m [27–29]. Such a value is comparable to that of the Bi/Ag interface [19]. It means that such an interfacial spin splitting is expected to lead to a sizable REMR in metallic bilayers, which is suggested by both theory [30] and experiment [31]. Because of this, in Pt/ferromagnet bilayers, the measured MR signals following the geometry of SMR may result from a concerted action of charge-spin inter-conversions that stem from both the bulk and the hetero-interface [32–44]. Nevertheless, up to now, a corresponding quantitative experimental analysis has remained lacking.

In this work, we quantitatively analyze the spin-current-related MR in epitaxial Pt/Co bilayers, by taking into account the charge-spin inter-conversions originating from both SHE and REE. By using a drift-diffusion model that incorporates both effects, the experimentally obtained thickness-dependent MR data can be well fitted, yielding a temperature-independent Rashba parameter that is consistent with theoretical calculations, along with an effective REE thickness that agrees with our previous spin-orbit torque analysis.

## II. EXPERIMENTAL METHODS

Trilayer thin films of Pt ($t_N$)/Co ($t_F$)/$AlO_x$ (2) (thickness in nanometer hereafter) were deposited onto $Al_2O_3$ (0001) substrates at room temperature using an ultrahigh vacuum magnetron sputtering facility. The $t_N$ and $t_F$ denote the thicknesses of Pt and Co, and are in the ranges of 0.2–5 nm and 0.6–10 nm, respectively. Wedged thin films were prepared by using a linear shutter that was moving at a constant speed of 0.1 mm/s during deposition. The (111)-oriented epitaxial growth of Pt/Co bilayers on $Al_2O_3$ (0001) was confirmed by structural characterizations such as x-ray diffraction, reflection high-energy electron diffraction and x-ray reflectivity. Details of structural characterizations can be found in a previous study [38]. In addition to epitaxial samples, polycrystalline thin films of Pt ($t_N$)/Co (3)/$AlO_x$ (2) were prepared on Si/$SiO_x$ substrates as control samples. The interfacial roughness at Pt/Co interface was below 0.1 nm for epitaxial samples and above 0.3 nm for polycrystalline samples. To measure the spin-current-related MR of Pt/Co bilayers, wedged thin films were fabricated into Hall bar devices that are 10 μm wide and 20-160 μm long using photo-lithography and Ar ion milling. Flat-film Hall bar devices with fixed Pt layer thickness were used to calibrate the Pt thickness of wedged Hall bar devices (see Supplemental Material [45] for details).

The definition of coordinate axis and the geometry of the Hall bar device for magneto-transport measurement are shown in Fig. 1(a). With an electric current $I_{xx}$ applied in the $x$ direction, the longitudinal voltage drop $V_{xx}$ is measured. The intrinsic spin-current-related MR ratio, following the geometry of the SMR, is defined as (see Supplemental Material [45] for a derivation using the parallel-circuit model):

$$MR_{\text{int}} \equiv \frac{\Delta R_{xx}^{y-z}}{R_{xx}^{0}} \approx \frac{R_{xx}(H_y) - R_{xx}(H_z)}{R_{xx}(H_z)} - MR_{\text{paras}}$$

where $R_{xx}(H_y)$ and $R_{xx}(H_z)$ refer to the longitudinal resistance $R_{xx}$ of the Hall bar device when Co magnetization is saturated along the $y$ and $z$ direction with an external magnetic field of 3 T. $MR_{\text{paras}}$ refers to the spin-current-unrelated parasitic MR arising from the Co layer itself due to the geometrical size effect [46], and it was obtained by performing additional MR measurement for a single Co layer following the same measurement geometry (see Supplemental Material [45] for details).

## III. CONVENTIONAL SMR MODEL ANALYSIS

As an initial step, the MR results of polycrystalline and epitaxial Pt ($t_{\text{N}}$)/Co (3) bilayers are analyzed and compared by using the conventional SMR model. Fig. 1 (b) and (c) show typical $R_{xx}$ plotted as a function of external magnetic field for polycrystalline Pt (1)/Co (3) and epitaxial Pt (1.8)/Co (3) bilayers. Note that the large $y$-$z$ resistance difference [solid circles and triangles in Fig. 1(b) and (c)] for both samples indicates strong charge-spin inter-conversions in Pt/Co bilayers. Fig. 1(d) and (e) show the room-temperature inverse sheet resistance and the spin-current-related MR ($\Delta R_{xx}^{y-z}/R_{xx}^{0}$) for Pt ($t_{\text{N}}$)/Co (3) bilayers. For a quantitative analysis of the experimental results, the drift-diffusion model, which has been proved effective to describe the spin-current transport [9,30,33,35], is utilized in this work. Following the SMR theory [9], the following equation is used to fit the data in Fig. 1(e):

$$\frac{\Delta R_{xx}^{\text{SMR}}}{R_{xx}^{0}} = -\theta_{\text{SH}}^{2} \frac{\lambda_{\text{sf}}}{t_{\text{N}}} \frac{\tanh(t_{\text{N}}/2\lambda_{\text{sf}})}{1+\eta} \left[1 - \frac{1}{\cosh(t_{\text{N}}/\lambda_{\text{sf}})}\right] \tag{1}$$

where $\theta_{\text{SH}}$ and $\lambda_{\text{sf}}$ are the spin Hall angle and spin diffusion length in Pt. $\eta = \rho_{\text{N}} t_{\text{F}}/\rho_{\text{F}} t_{\text{N}}$ is the current shunting coefficient, where $\rho_{\text{N}}$ and $\rho_{\text{F}}$ are the resistivity of Pt and Co. For simplicity, we assume a transparent Pt/Co interface similar to previous SMR studies on Pt/Co bilayers [20,21]. Also, the MR contribution from the longitudinal spin absorption term [11] is neglected in Eq. (1) because such a term is negligibly small due to the long spin diffusion length in Co [47].

As a result, both polycrystalline and epitaxial data at 300 K can be reasonably fitted as seen in Fig. 1(e). For polycrystalline samples, $\theta_{\text{SH}}$ of 0.19 and $\lambda_{\text{sf}}$ of 0.93 nm are obtained, which is in overall agreement with earlier studies [48–50], suggesting that the SMR model may be applicable to the analysis of spin transport in polycrystalline samples. For epitaxial samples, surprisingly, a large $\theta_{\text{SH}}$ of 0.33 and a short $\lambda_{\mathrm{sf}}$ of 0.35 nm are obtained from the curve fitting. In the following, we discuss the values of the extracted $\lambda_{\mathrm{sf}}$ and $\theta_{\mathrm{SH}}$ from SMR analysis in epitaxial samples. Firstly, according to previous studies, the experimentally obtained values of $\lambda_{\mathrm{sf}}$ range from 1–10 nm [1,38,48–55]. In particular, a previous current-perpendicular-to-plane giant magnetoresistance study reported a $\lambda_{\mathrm{sf}}$ of about 3.5 nm at 300 K [54]. Based on this, apparently, the extracted $\lambda_{\mathrm{sf}}$ of 0.35 nm via SMR fitting is much smaller than its lower-bound values and is probably underestimated. Secondly, previous studies have shown that the $\theta_{\mathrm{SH}}$ of pure Pt, scaling linearly with its resistivity [38,52], is typically less than 0.2 [1,38,48–52]. For instance, our previous harmonic Hall analysis for the same epitaxial Pt/Co bilayers yields a $\theta_{\mathrm{SH}}$ of 0.03 at 300 K [38]. Therefore, the extracted $\theta_{\mathrm{SH}}$ of 0.33, being about one order of magnitude greater, is probably an overestimated value. Based on the discussion above, we conclude that the SMR model alone is *invalid* for analyzing the spin-current-related MR in our epitaxial Pt/Co bilayers. On the other hand, it is suggested from the discussion above that the observed MR originates from charge-spin inter-conversions that take place within a much shorter length scale, *i.e.* the MR is probably dominated by REE instead of SHE. Therefore, in the following, we analyze the spin-current-related MR data in epitaxial samples by using a combined SHE-REE MR model where both effects are taken into account.

## IV. COMBINED SHE-REE MR MODEL ANALYSIS

A combined SHE-REE MR model is derived by taking into account both effects in the spin-current generation. Here, a drift-diffusion model is developed by adding an Edelstein contribution of spin accumulation [35] into the SMR theory [9]. The combined SHE-REE MR reads [56] (see Supplemental Material [45] for a detailed derivation):

$$\frac{\Delta R_{xx}}{R_0} \approx -\frac{\lambda_{\mathrm{sf}}}{(1+\eta)t_{\mathrm{N}}}\left[\frac{g_R}{1+g_R\coth(t_{\mathrm{N}}/\lambda_{\mathrm{sf}})}\right](\Omega_{\mathrm{SHE}}+\Omega_{\mathrm{REE}}+\Omega_{\mathrm{mix}}) \tag{2}$$

Here,

$$g_R \equiv 2\rho_{\mathrm{N}}\lambda_{\mathrm{sf}}\mathrm{Re}[G_{\uparrow\downarrow}]$$

$$\Omega_{\mathrm{SHE}} \equiv \theta_{\mathrm{SH}}^2\tanh^2\left(\frac{t_{\mathrm{N}}}{2\lambda_{\mathrm{sf}}}\right)$$

$$\Omega_{\mathrm{REE}} \equiv \left(\frac{\lambda_{\mathrm{IEE}}}{2\lambda_{\mathrm{sf}}}\right)^2 \left\{1 - \frac{2\langle\sinh[(t_{\mathrm{N}} - d_2)/\lambda_{\mathrm{sf}}]\rangle}{\sinh(t_{\mathrm{N}}/\lambda_{\mathrm{sf}})} + \frac{\langle\sinh^2[(t_{\mathrm{N}} - d_2)/\lambda_{\mathrm{sf}}]\rangle}{\sinh^2(t_{\mathrm{N}}/\lambda_{\mathrm{sf}})}\right\}$$

$$\Omega_{\mathrm{mix}} \equiv 2\theta_{\mathrm{SH}} \left(\frac{\lambda_{\mathrm{IEE}}}{2\lambda_{\mathrm{sf}}}\right) \left\{1 - \frac{\langle\sinh[(t_{\mathrm{N}} - d_2)/\lambda_{\mathrm{sf}}]\rangle}{\sinh(t_{\mathrm{N}}/\lambda_{\mathrm{sf}})}\right\} \tanh\frac{t_{\mathrm{N}}}{2\lambda_{\mathrm{sf}}}$$

where $\Omega_{\mathrm{SHE}}$, $\Omega_{\mathrm{REE}}$ and $\Omega_{\mathrm{MIX}}$ are the MR components induced by the SHE term, the REE term and the SHE-REE cross term, respectively. $\lambda_{\mathrm{IEE}}$ is the inverse REE length [18,57] which characterizes the efficiency of the interfacial charge-spin conversion [18,58,59], $d_2$ is the thickness of the RE region, $G_{\uparrow\downarrow}$ is the spin-mixing conductance at the Pt/Co interface [60], $\langle\cdots\rangle$ denotes the average with respect to $d_2$ weighted by $\exp(-d_2/d_{\mathrm{R}})$, and $d_{\mathrm{R}}$ is the effective REE thickness of an equivalent layer where the spin accumulation density is homogenous.

As for the SHE-REE MR model, the fitting parameters are $\lambda_{\mathrm{IEE}}$ and $d_{\mathrm{R}}$. The details of the nonlinear-fitting procedure are described below: firstly, a transparent Pt/Co interface ($\mathrm{Re}[G_{\uparrow\downarrow}] \to \infty$) is assumed which is the same as that in the SMR analysis; secondly, $\theta_{\mathrm{SH}}$ and $\lambda_{\mathrm{sf}}$ are set as fitting constant, which are directly obtained from a previous experimental spin-orbit torque study [38] for the same epitaxial Pt/Co bilayers (Table. 1). It is also worth to note that $\theta_{\mathrm{SH}}$ and $\lambda_{\mathrm{sf}}$ shown in Table. 1 suggest the Elliott-Yafet spin relaxation [61,62] and the intrinsic and/or side-jump contribution to the SHE [1], which is consistent with previous spin transport studies for Pt [38,52,63].

Figure 2(a) shows the MR analysis by using the SHE-REE MR model for Pt/Co (3) bilayers at 300 K. The experimental data are well fitted by this new model. To acquire a better understanding of the fitting output, the fitting curve is decomposed into three components: the SHE term, the REE term and the SHE-REE cross term. As a result, the fitting curve is dominated by the REE term in view of the magnitude of the MR signal, whereas the SHE term is almost negligible. This is consistent with the conclusion in Section III that the observed MR signal in epitaxial Pt/Co bilayers is dominated by the *interfacial* charge-spin inter-conversions. It is also worth to note that the peak position of the fitting curve is determined by both the REE term and the SHE-REE cross term.

Figure 2(b)–(f) show the spin-current-related MR ($\Delta R_{xx}^{y-z}/R_{xx}^{0}$) data plotted as function of Pt layer thickness $t_{\mathrm{N}}$ at different temperatures and the corresponding fitting curves by using the SHE-REE MR model. All the experimental data are well fitted for $t_{\mathrm{F}} \geq 1$ nm. In the case of $t_{\mathrm{F}} = 0.6$ nm, the deviation of fitting curves from the MR data implies that the fitting equation may no longer be suitable to describe the bilayer spin transport for very small $t_{\mathrm{F}}$. Such a deviation is likely to arise from the effect of spin backflow at Co/AlO$_x$ interface because such a $t_{\mathrm{F}}$ is smaller than the transverse spin absorption length of about 1.2 nm in Co [64]. Note that such a spin backflow effect on the spin-current transport was also discussed in a previous spin-orbit torque study for Pt-based multilayer [65] thin films, where a dramatic modulation of spin current was observed. We also suggest that such an insufficient transverse spin absorption may account for the reversed temperature dependence of the MR data shown in Fig. 2(b), that for a given $t_{\mathrm{N}}$, the overall MR decreases with decreasing temperature, whereas such a temperature dependence is reversed for larger $t_{\mathrm{F}}$ as shown in Fig. 2(d)-2(f).

The SHE-REE MR fitting results are summarized in Fig. 3. As shown in Fig. 3(a), for a fixed temperature, $\lambda_{\mathrm{IEE}}$ gradually increases with increasing $t_{\mathrm{F}}$ and saturates for $t_{\mathrm{F}} \geq 6$ nm. Such $t_{\mathrm{F}}$ thickness dependences could also be related to the spin backflow effect at Co/AlO$_x$ interface. The reason is as follows: according to a previous study [65], the spin backflow effect is expected to suppress the overall charge-spin conversion when $t_{\mathrm{F}}$ is smaller than the Co transverse spin absorption length (1.2 nm) [64,65]. Because such an effect becomes weaker for larger $t_{\mathrm{F}}$, the overall charge-spin conversion efficiency is expected to increase with $t_{\mathrm{F}}$ and saturate as $t_{\mathrm{F}}$ becomes much greater than 1.2 nm, which is consistent with the results shown in Fig. 3(a). Thus, the $t_{\mathrm{F}}$ thickness dependence of $\lambda_{\mathrm{IEE}}$ can be qualitatively described by such a spin backflow picture. In addition to the $t_{\mathrm{F}}$ dependence, $\lambda_{\mathrm{IEE}}$ also depends on temperature. On the other hand, such a $t_{\mathrm{F}}$ and temperature dependences are not systematically observed for $d_{\mathrm{R}}$ shown in Fig. 3(b).

## V. DISCUSSION

### A. $d_{\mathrm{R}}$ and $\lambda_{\mathrm{IEE}}$

We have quantitatively analyzed the spin-current-related MR ($\Delta R_{xx}^{y-z}/R_{xx}^{0}$) in epitaxial Pt/Co bilayers by using a drift-diffusion model that takes into account both the SHE and the REE. Next we discuss the fitting results of $d_{\mathrm{R}}$ and $\lambda_{\mathrm{IEE}}$. As shown in Fig. 3(b), no systematic $t_{\mathrm{F}}$ or temperature dependences are observed for $d_{\mathrm{R}}$, indicating that the maximum charge-spin inter-conversion is achieved with a similar Pt thickness. Accordingly, a zero-slope linear fitting is performed for $d_{\mathrm{R}}$ shown in Fig.

3(b), where the *y*-axis intercept of the fitting lines corresponds to an average value (*i.e.* $d_{\mathrm{R0}}$). The values of $d_{\mathrm{R0}}$ are summarized in Fig. 4(b) (solid circles). As expected, $d_{\mathrm{R0}}$ is almost temperature-independent. Considering that the REE is strongly associated with the electron screening effect by an interfacial electric field, our results suggest that the out-of-plane penetration depth of Pt/Co interfacial electric field (*e.g.* towards the Pt side) may not be significantly affected by thermal agitations such as that by phonons in the temperature range of 10–300 K. The extracted $d_{\mathrm{R0}}$ in this work is in overall consistency with values extracted from our previous spin-orbit torque study for the same epitaxial Pt/Co bilayers [38], which are also shown in Fig. 4(b) for a comparison.

In contrast to $d_{\mathrm{R}}$, the $\lambda_{\mathrm{IEE}}$ shows systematic temperature dependence. For a quantitative analysis, a zero-slope linear fitting is performed for the experimental data shown in Fig. 3(a) in the range of $t_{\mathrm{F}} \geq 3$ nm (dashed lines), where bulk-interface charge-spin conversion is expected to saturate because such a $t_{\mathrm{F}}$ range is much greater than the Co transverse spin absorption length (1.2 nm) [64]. The *y*-axis intercept of the fitting lines corresponds to a saturated value of $\lambda_{\mathrm{IEE}}$ ($\lambda_{\mathrm{IEE0}}$). As a result, the $\lambda_{\mathrm{IEE0}}$ increases from 1.5 nm to 3.5 nm when temperature drops from 300 K to 10 K. A brief discussion on the magnitude of $\lambda_{\mathrm{IEE0}}$ and REMR suggests that these $\lambda_{\mathrm{IEE0}}$ values are plausible for our epitaxial Pt/Co interface (see Supplemental Material [45]).

### B. Conversion of $\lambda_{\mathrm{IEE}}$ to the Rashba parameter

To acquire a physical interpretation of $\lambda_{\mathrm{IEE0}}$, it is converted to the Rashba parameter $\alpha_{\mathrm{R}}$ via $\alpha_{\mathrm{R}} = \hbar\lambda_{\mathrm{IEE0}}/\tau_{\mathrm{p}}$ [18,57], where $\tau_{\mathrm{p}}$ is the momentum scattering time in Pt and $\hbar$ the Dirac constant. By plugging a carrier density of $10^{28}$ $\mathrm{m}^{-3}$ in Pt [66] and assuming that the effective mass of the electron equals the free electron mass, the resultant $\alpha_{\mathrm{R}}$ turns out to be almost temperature-independent [Fig. 4(a)], where an average value of $(4.1\pm0.2)\times10^{-11}$ eV m is obtained. Such a value is, to some extent, consistent with previous theoretical calculations of $\alpha_{\mathrm{R}}$ at Pt interfaces (e.g. $10\times10^{-11}$ eV m in Ref. [27] and $2.5\times10^{-11}$ eV m in Ref. [28]).

## VI. OUTLOOK

We compare our results in Table. 2 with previous experimental studies on various Rashba systems, *e.g.* Bi/Ag [13,18,67–69], Bi/Cu [70], $Bi_2O_3$/Cu [71] and $SrTiO_3$/$LaAlO_3$ (STO/LAO) [58,72] interfaces. Among all the systems, the Bi/Ag interface exhibits one of the largest $\alpha_{\mathrm{R}}$ [13], which is consistent with the angle-resolved photoelectron spectroscopy results revealing a giant interfacial spin splitting [19]. In spite of this, the largest $\lambda_{\mathrm{IEE0}}$ is found in STO/LAO two-dimensional electron gas, with $\alpha_{\mathrm{R}}$ being one to two orders of magnitude smaller. This is mainly due to a much larger $\tau_{\mathrm{p}}$ in STO/LAO Rashba systems because of their ultralow sheet carrier density. Actually, even greater $\lambda_{\mathrm{IEE0}}$ of 20−60 nm have been demonstrated in STO/LAO system [73,74]. We suggest here that a much enhanced $\lambda_{\mathrm{IEE0}}$ might be obtained in epitaxial Pt/Co bilayers via a carrier density (or carrier life time) engineering [75].

Finally, we briefly note that, as the Pt layer thickness $t_{\mathrm{N}}$ in this study is comparable to (or even smaller than) its spin diffusion length, the actual electronic transport may go beyond the drift-diffusion model from a more critical point of view. The actual electronic transport in Pt is very likely to be in an intermediate regime that neither ballistic nor diffusive approximation *alone* is valid to describe. Future studies are required to present a more precise transport model to account for the spin-related MR in nanoscale heavy-metal/ferromagnet bilayers.

## VII. CONCLUSION

In conclusion, we have analyzed the experimentally obtained spin-current-related magnetoresistance in epitaxial Pt/Co bilayers by using a drift-diffusion model that takes into account both bulk SHE and interfacial REE. The magnetoresistance analysis yields, for the Pt/Co interface, a reasonable Rashba parameter in the order of $10^{-11}$ eV m and an effective REE thickness of several angstroms, both of which are almost temperature independent. Our results show that both bulk SHE and interfacial REE have to be considered as the source of spin-current generation in order to properly quantify the spin-current-related MR in epitaxial Pt/Co bilayers.

## Acknowledgements

We would like to thank Prof. Makoto Kohda, Shutaro Karube, Jeongchun Ryu, Hiromu Gamou and Ryan Thompson for fruitful discussions. This work was partially supported by the Japanese Ministry of Education, Culture, Sports, Science, and Technology (MEXT) in Grant-in-Aid for Scientific Research (Grant No. 15H05699) and the JSPS Core-to-Core program (Grant No. JPJSCCA20160005). Y. D. acknowledges the Center for Science and Innovation in Spintronics (CSIS), Tohoku University for financial support.

## Corresponding author:

*Author to whom correspondence should be addressed.

E-mail address:

nitta@material.tohoku.ac.jp

**TABLE 1**. Temperature dependence of resistivity, spin-diffusion length and spin Hall angle of epitaxial Pt obtained from our previous spin-orbit torque analysis for the same epitaxial Pt/Co bilayers (data taken from Ref. [38]).

| Temperature (K) | Pt resistivity ($\mu\Omega$ cm) | Spin diffusion length (nm) | Spin Hall angle |
|---|---|---|---|
| 300 | 14.87 | 1.75 | 0.029 |
| 200 | 12.56 | 1.78 | 0.022 |
| 100 | 10.12 | 2.30 | 0.018 |
| 4.2 | 7.29 | 3.46 | 0.013 |

**TABLE 2**. A summary of the absolute values of $\alpha_{\mathrm{R}}$, $\lambda_{\mathrm{IEE}}$ and $\tau_{\mathrm{p}}/\hbar$ obtained from previous studies and current work for different Rashba systems.

| Hetero-interface | $\|\boldsymbol{\alpha}_{\mathbf{R}}\|$ (eV Å) | $\boldsymbol{\tau}_{\mathbf{p}}/\hbar$ ($eV^{-1}$) | $\|\boldsymbol{\lambda}_{\mathbf{IEE}}\|$ (nm) | Reference |
|---|---|---|---|---|
| Bi/Ag | 0.56 | 5.4 | 0.3 | [18] |
| Bi/Ag | 1.7 | 1.8 | 0.3 | [13] |
| Bi/Cu | 0.56 | 0.16 | 0.009 | [70] |
| $Bi_2O_3$/Cu | 0.46 | 13 | 0.6 | [71] |
| $SrTiO_3$/$LaAlO_3$ | 0.030 | 2133 | 6.4 | [58] |
| $SrTiO_3$/$LaAlO_3$ | 0.0035 | 19143 | 6.7 | [72] |
| Pt/Co | 0.41 | 37−85 | 1.5−3.5 | This work |

## Figures and figure captions

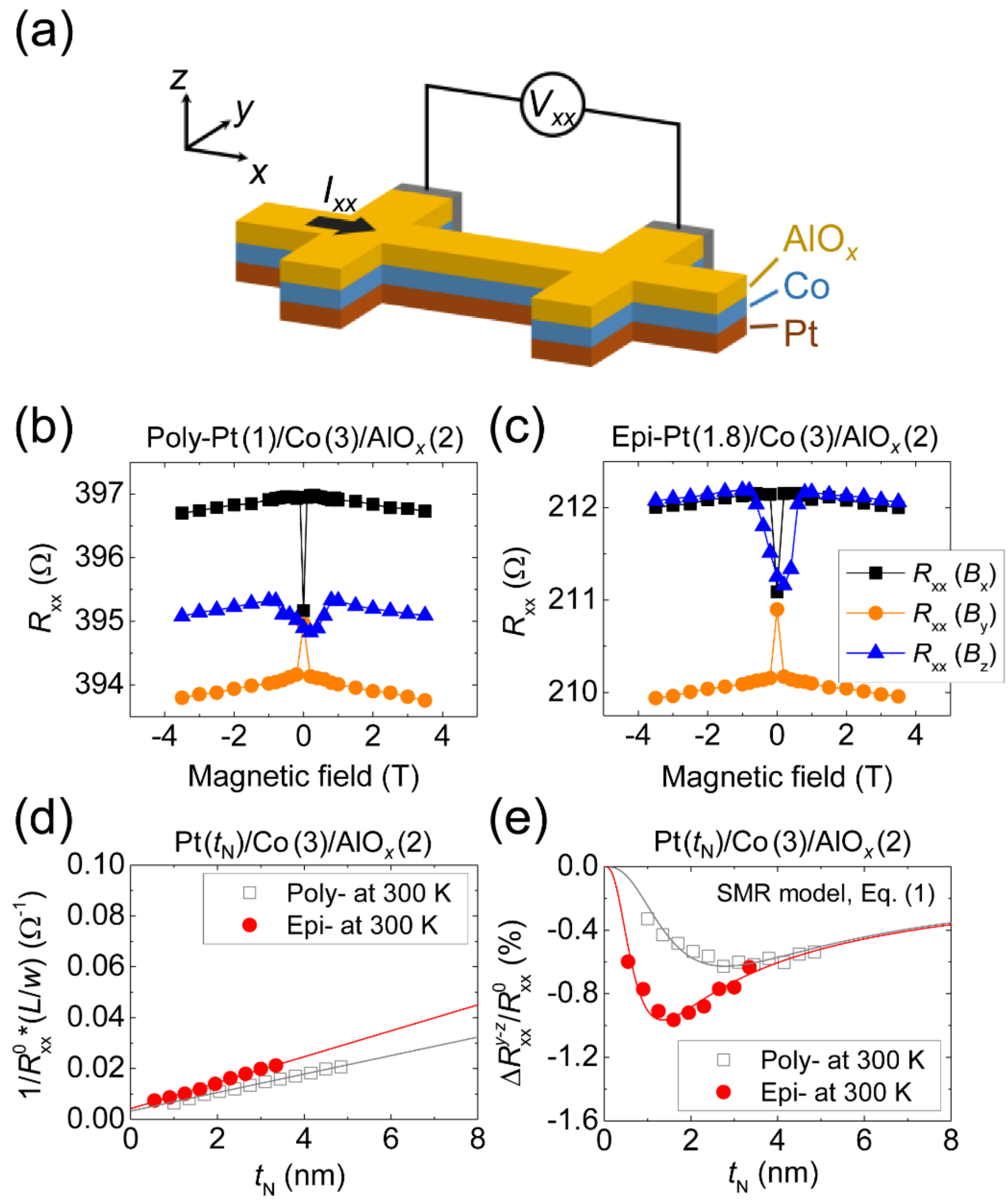

**Fig. 1** (a) Definition of coordinate axis and schematic demonstration of the geometry of the Hall bar device for magneto-transport measurement. (b), (c) Longitudinal resistance $R_{xx}$ of polycrystalline Pt (1)/Co (3)/AlO$_x$ (2) and epitaxial Pt (1.8)/Co (3)/AlO$_x$ (2) Hall bar devices with external magnetic field applied in the $x$, $y$ and $z$ direction. (d), (e) Pt thickness $t_{\mathrm{N}}$ dependence of inverse sheet resistance (d) and spin-current-related MR (e) for both polycrystalline and epitaxial samples. Solid lines in (d) are linear fittings to the data; solid lines in (e) are nonlinear fittings via Eq. (1) with the SMR model.

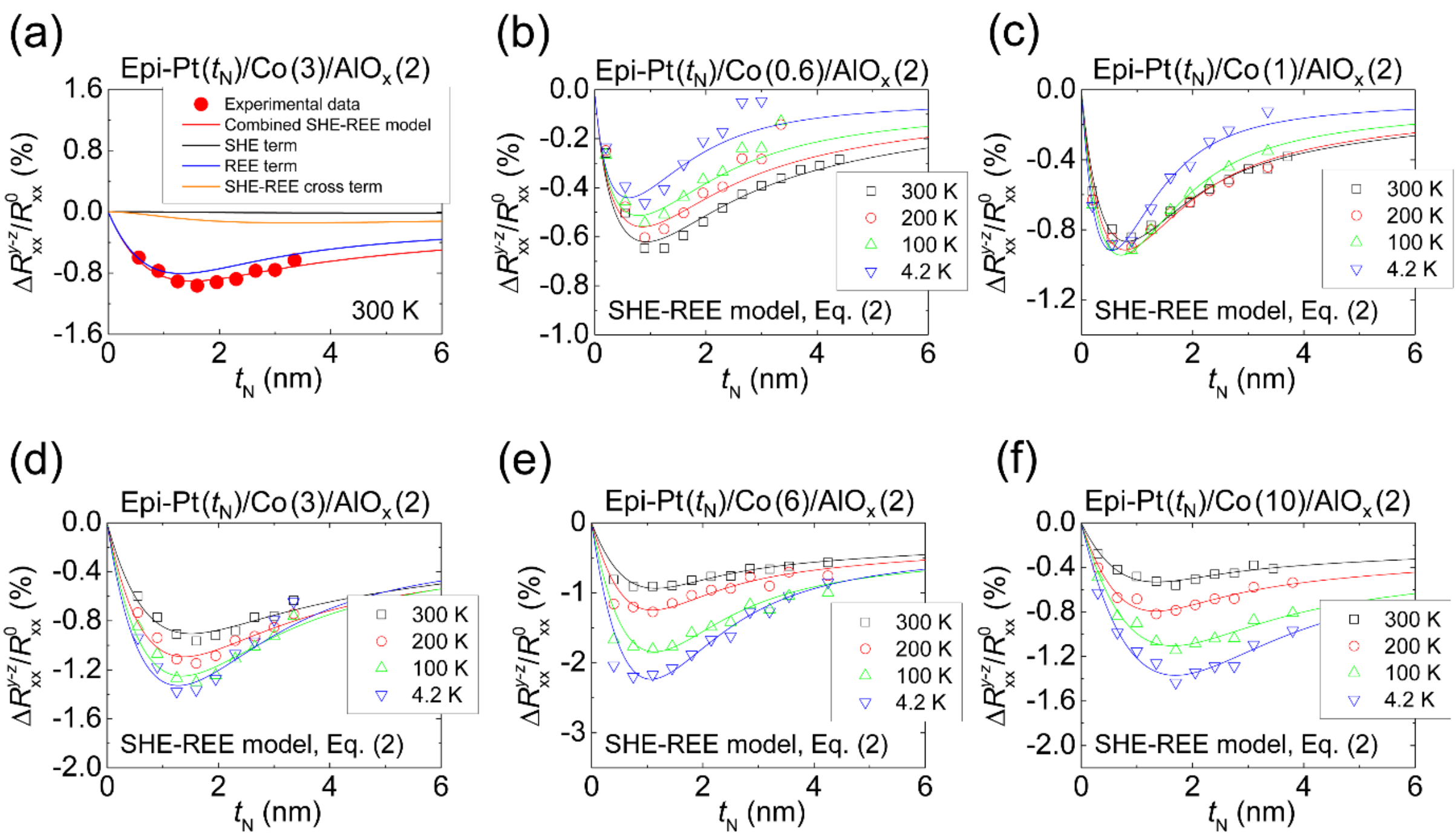

**Fig. 2** (a) The combined SHE-REE MR model analysis via Eq. (2) for experimental data of epitaxial Pt/Co (3) bilayers at 300 K. (b)−(e) Pt layer thickness $t_N$ and temperature dependence of the spin-current-related MR data for epitaxial Pt/Co bilayers with $t_F$ = 0.6 nm (b), $t_F$ = 1.0 nm (c), $t_F$ = 3.0 nm (d), $t_F$ = 6.0 nm (e) and $t_F$ = 10.0 nm (f). Solid curves are fittings to the experimental data using Eq. (2) by plugging in experimentally extracted spin diffusion length and spin Hall angle shown in Table. 1.

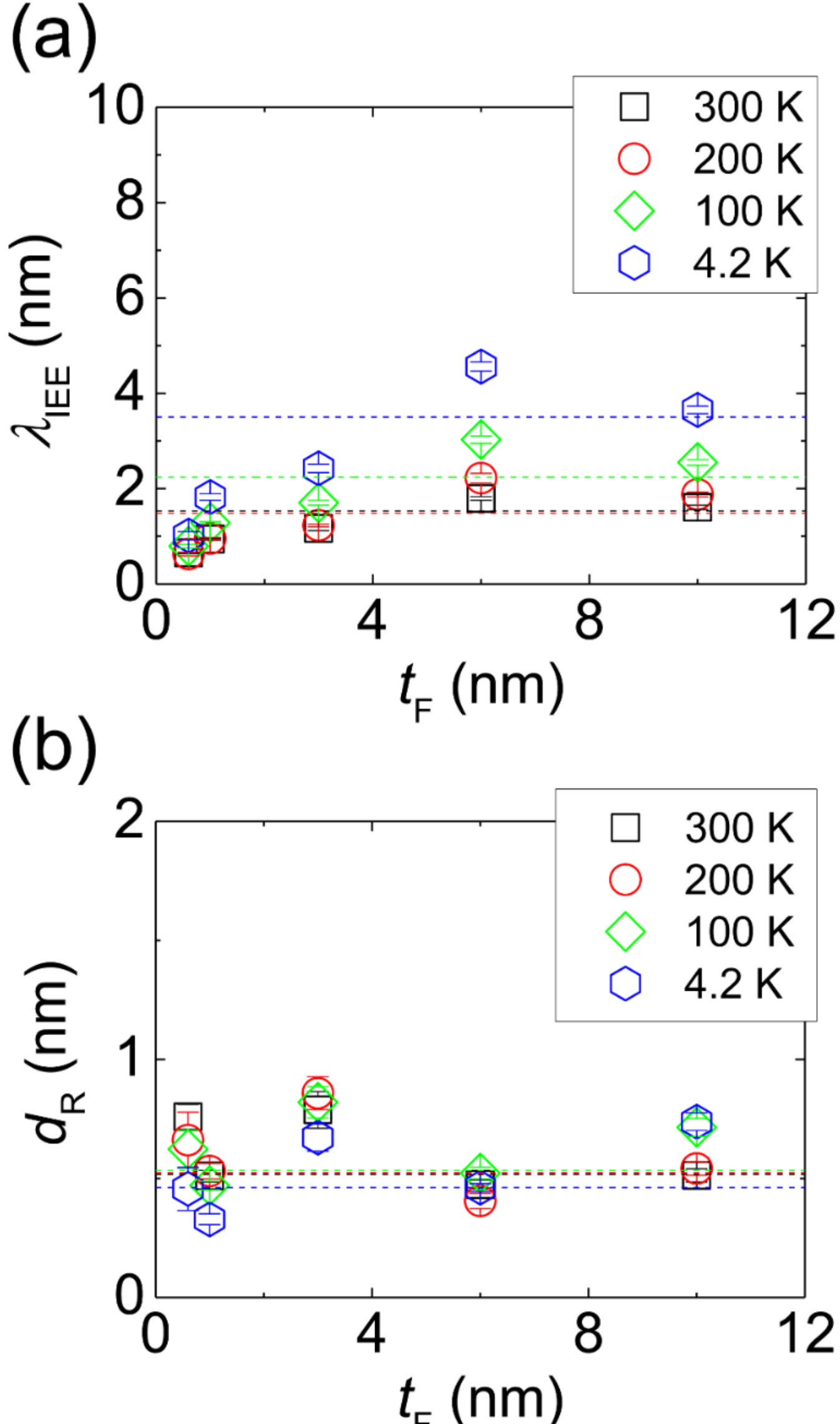

**Fig. 3** (a), (b) Experimentally extracted inverse REE length $\lambda_{IEE}$ (a) and effective REE thickness $d_R$ (b) at different temperatures plotted as a function of Co layer thickness $t_F$. Dashed lines in (a) are zero-slope linear fittings for $t_F$ ranging from 3-10 nm; in (b) are zero-slope linear fittings with full $t_F$ range.

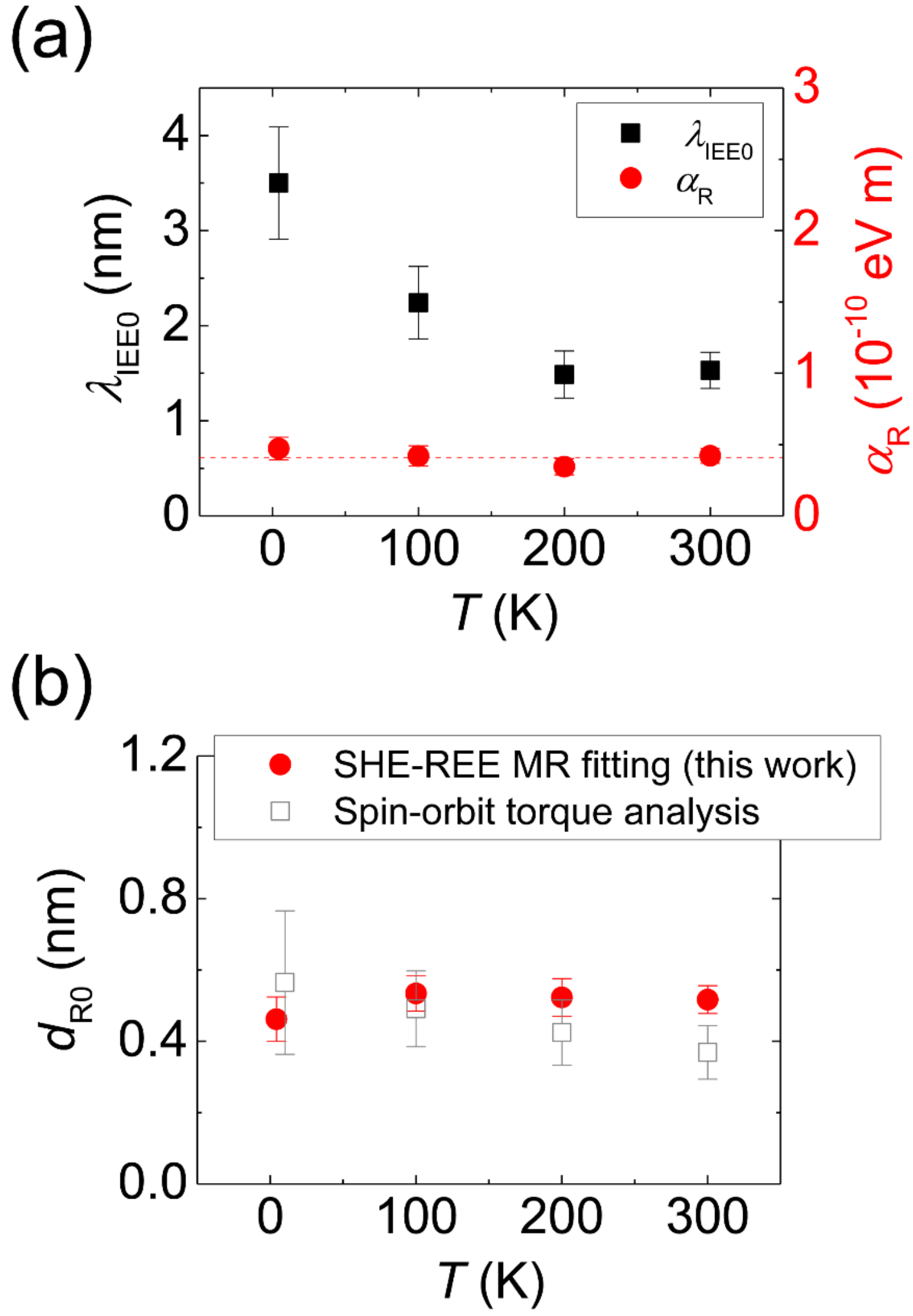

**Fig. 4** (a) The Rashba parameter $\alpha_R$ at Pt/Co interface (solid circles) and the saturated value of the inverse REE length $\lambda_{IEE0}$ (solid squares) plotted as a function of measurement temperature. The dashed line is a zero-slope linear fit for $\alpha_R$. (b) An average value of the effective REE thickness $d_{R0}$ (solid circles) plotted against measurement temperature; for a comparison, $d_{R0}$ obtained from a previous spin-orbit torque study from Ref. [38] are also shown (open squares).

Supplemental material for

# Spin-current-related magnetoresistance in epitaxial Pt/Co bilayers in the presence of spin Hall effect and Rashba-Edelstein effect

Ye Du[1,2], Saburo Takahashi[3], and Junsaku Nitta[1,2,4,*]

[1]Department of Materials Science, Tohoku University, Sendai 980-8579, Japan

[2]Center for Science and Innovation in Spintronics, Tohoku University, Sendai 980-8577, Japan

[3]Advanced Institute for Materials Research, Tohoku University, Sendai 980-8577

[4]Center for Spintronics Research Network, Tohoku University, Sendai 980-8577, Japan

**Supplementary Figure S1**

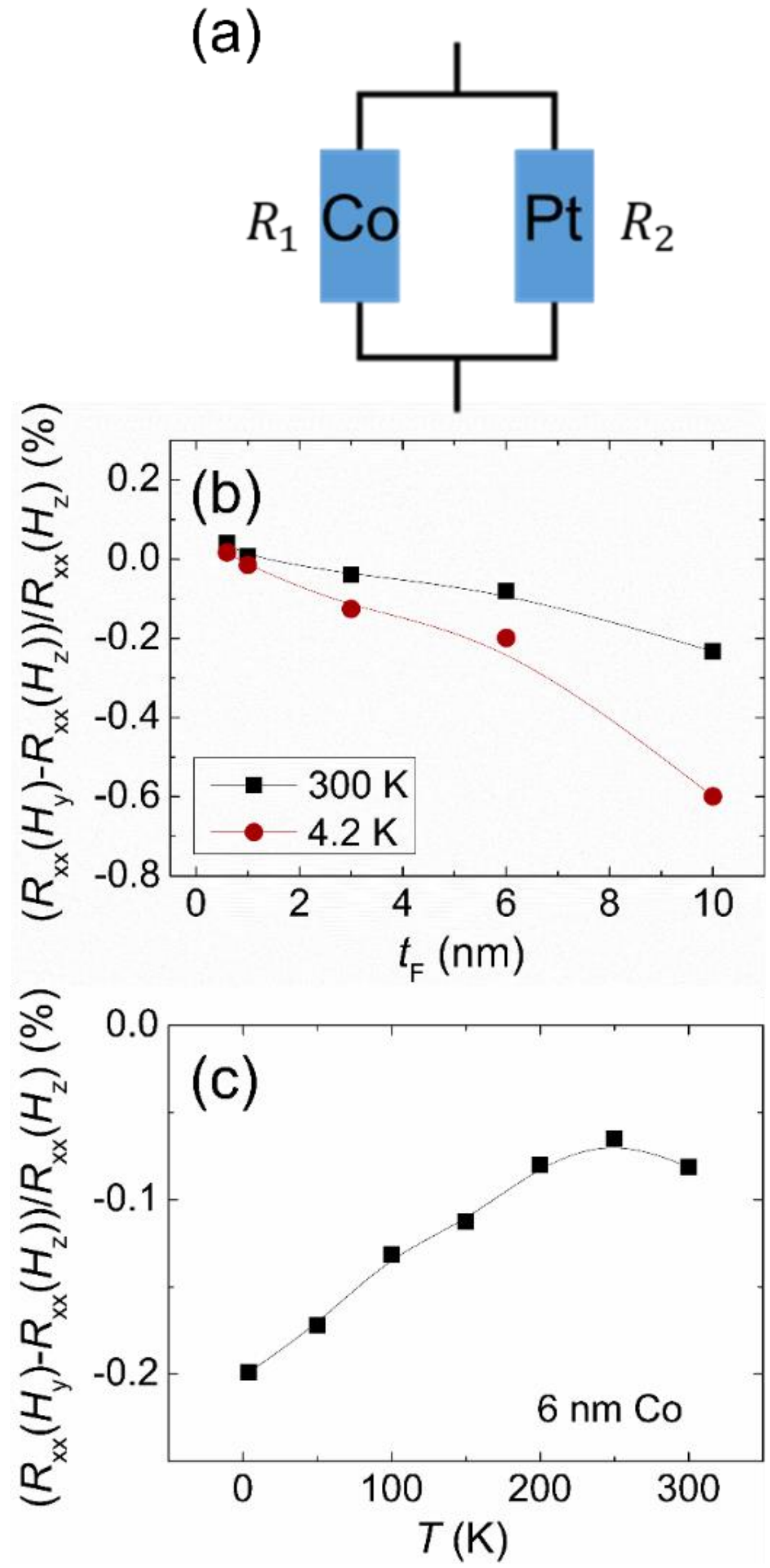

Supplementary Figure S1: **Calibration of the spin-Hall Rashba-Edelstein magnetoresistance.** (a) Schematic demonstration of the parallel configuration for Co and Pt conduction channels. (b) Parasitic MR in single Co layer at 300 K and 4.2 K. (c) Temperature dependence of parasitic MR for 6 nm Co layer.

**Supplementary Figure S2**

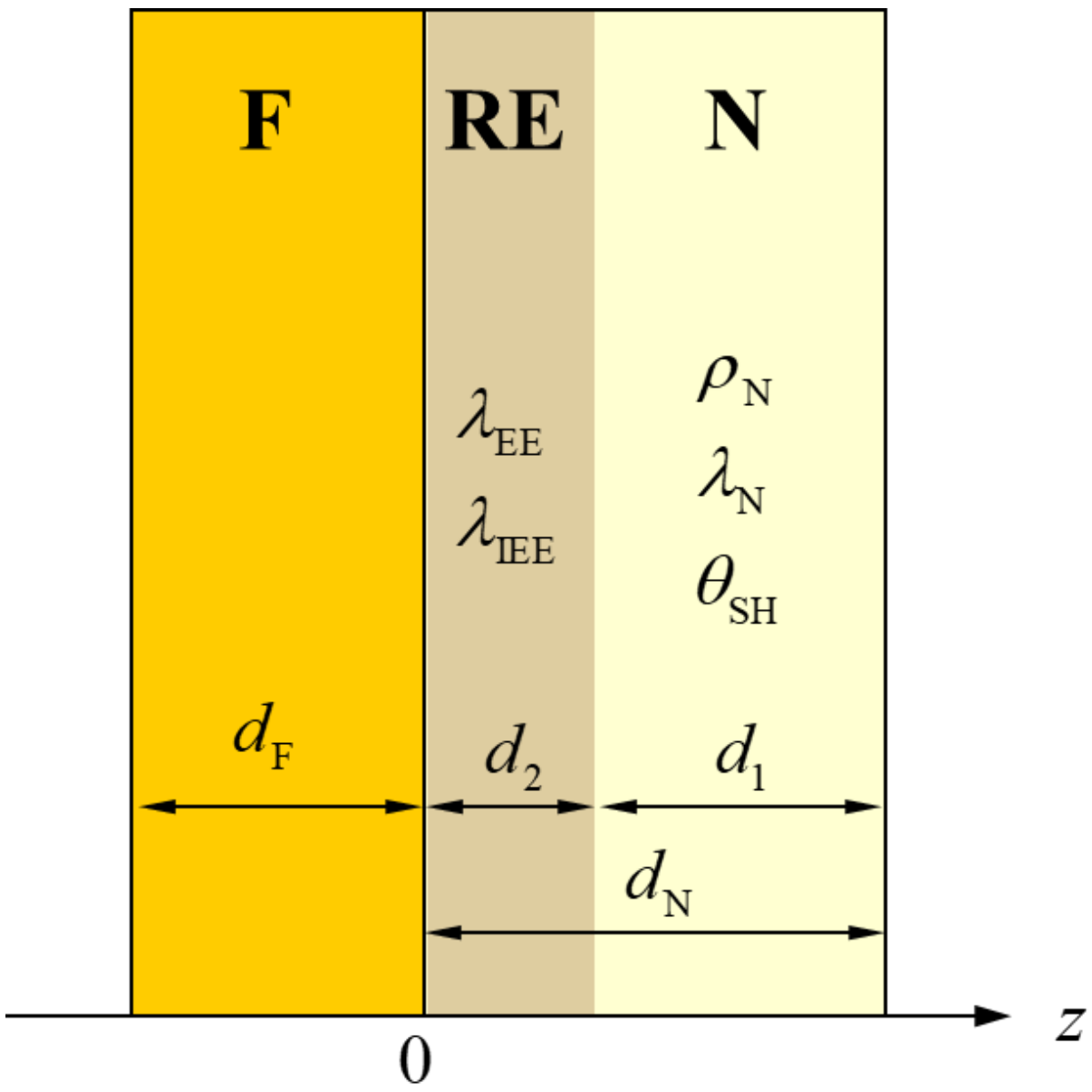

Supplementary Figure S2: **Derivation of the spin-Hall Rashba-Edelstein magnetoresistance.** Schematic demonstration of a hetero-structure used for deriving the spin-Hall Rashba-Edelstein MR. The Rashba-Edelstein (RE) region with a thickness $d_2$ is superimposed on the left side of non-magnetic layer (N) with a thickness of $d_N$. $\lambda_{IEE}$ and $\lambda_{EE}$ are the inverse and direct Edelstein length in the RE region. $\rho_N$, $\lambda_N$ and $\theta_{SH}$ are the resistivity, spin diffusion length and spin Hall angle in N. $\rho_F$ and $d_F$ are the resistivity and thickness of the ferromagnetic layer (F).

**Supplementary Figure S3**

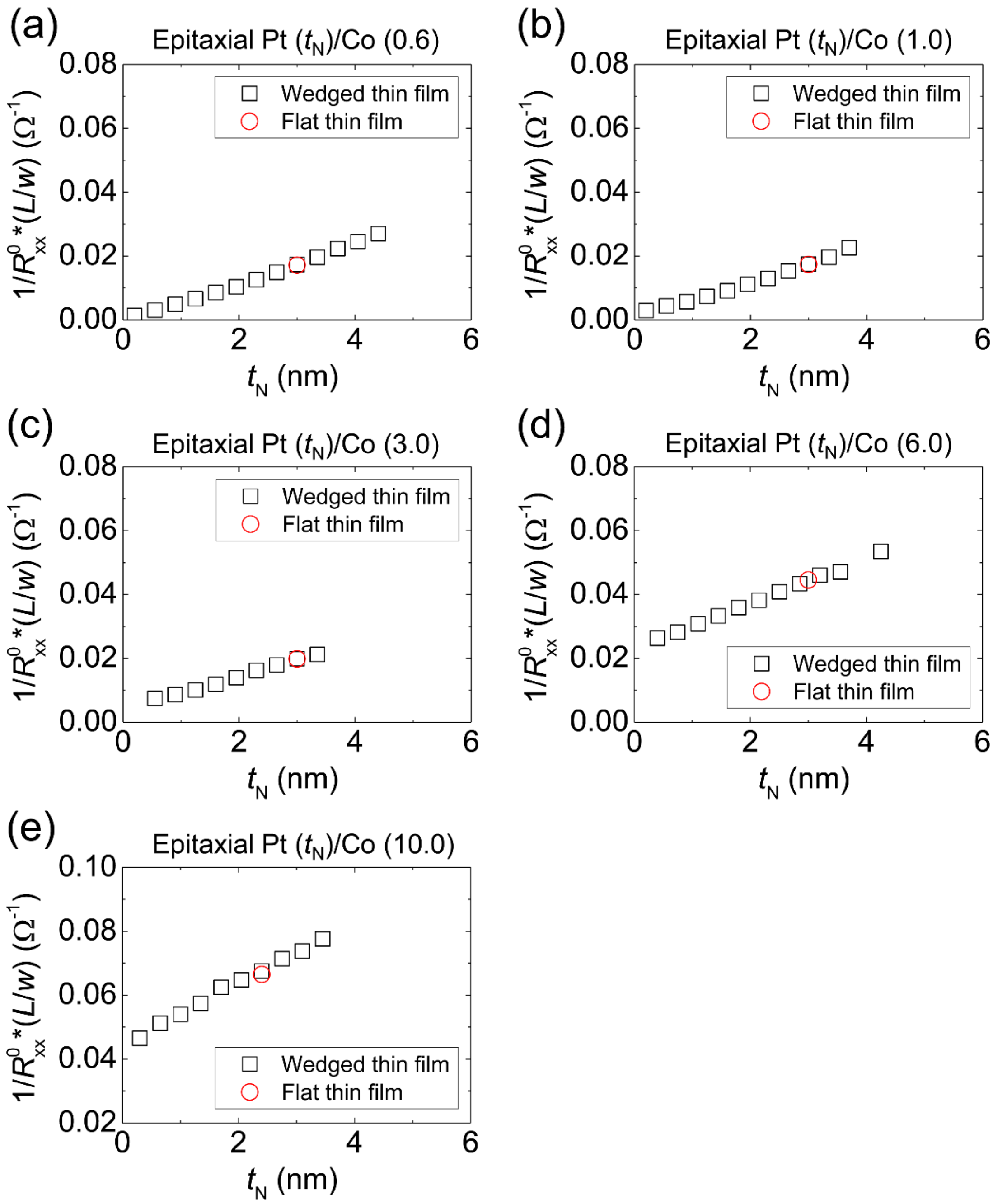

Supplementary Figure S3: **Thickness calibration for wedged thin films at 300 K.**

## Supplementary Note 1. Calibration of the spin-Hall Rashba-Edelstein magnetoresistance

Physical parameters such as the inverse Edelstein length ($\lambda_{\mathrm{IEE}}$), effective thickness of the Rashba region ($d_R$) *et al.* in this work are extracted from magnetoresistance (MR) that is originated from the charge-spin current interplay, *i.e.* the spin Hall effect (SHE) and the Rashba-Edelstein effect (REE). Therefore, parasitic MR originated from *e.g.* the geometrical size effect [1] has to be subtracted from the raw data to obtain genuine spin-current-related MR signal. To perform the MR calibration, we use a parallel circuit to describe the MR in the Pt/Co bilayer (Fig. S1). The longitudinal resistance with field applied in *z* and *y* direction are respectively

$$R_{xx}(H_z) = R_0 = \frac{R_1 R_2}{R_1 + R_2}$$

$$R_{xx}(H_y) = R_0' = \frac{R_1' R_2'}{R_1' + R_2'}$$

where $R_1' = R_1 + \Delta R_1$, $R_2' = R_2 + \Delta R_2$. $\Delta R_1$ is the parasitic resistance change in Co layer and $\Delta R_2$ is the *overall* resistance change induced by SHE and REE. Thus we have

$$R_{xx}(H_y) = \frac{(R_1 + \Delta R_1)(R_2 + \Delta R_2)}{R_1 + R_2 + \Delta R_1 + \Delta R_2} \approx R_0 + R_0(\frac{\Delta R_1}{R_1} + \frac{\Delta R_2}{R_2})$$

The MR originated from the SHE and REE ($MR_{int}$) and the raw MR ratio without calibration reads

$$MR_{int} \approx \frac{R_1(R_2 + \Delta R_2)}{R_1 + R_2 + \Delta R_2} - \frac{R_1 R_2}{R_1 + R_2} \approx \frac{\Delta R_2}{R_2}$$

$$MR_{raw} = \frac{R_{xx}(H_y) - R_{xx}(H_z)}{R_{xx}(H_z)} \approx \frac{\Delta R_1}{R_1} + \frac{\Delta R_2}{R_2} = MR_{int} + \frac{\Delta R_1}{R_1}$$

in which $\Delta R_1/R_1$ is the parasitic MR originated from single Co layer. Therefore, one obtains the $MR_{int}$ in Pt/Co bilayers by subtracting the parasitic MR from single Co layer with the same measurement geometry.

## Supplementary Note 2. Derivation of the spin-Hall Rashba-Edelstein magnetoresistance

The schematic model for the spin-Hall Rashba-Edelstein magnetoresistance is shown in Fig. S2. The Rashba-Edelstein (RE) region with a thickness $d_2$ is sandwiched by the ferromagnetic layer (F) with a thickness of $d_F$ and non-magnetic layer (N) with a thickness of $d_N$. By adding an Edelstein contribution $e(\lambda_{EE}\hat{\boldsymbol{z}} \times \boldsymbol{E})$ of spin accumulation [2] into the SMR theory [3], the spin accumulation in N reads:

$$\begin{cases} \delta\boldsymbol{\mu}_{N1}(z) = \boldsymbol{a}e^{-\frac{z-d_2}{\lambda_N}} + \boldsymbol{b}e^{\frac{z-d_2}{\lambda_N}}, & (d_2 < z < d_1 + d_2) \\ \delta\boldsymbol{\mu}_{N2}(z) = \boldsymbol{c}e^{-\frac{z}{\lambda_N}} + \boldsymbol{d}e^{\frac{z}{\lambda_N}} + e\lambda_{EE}\hat{\boldsymbol{z}} \times \boldsymbol{E}, & (0 < z < d_2) \end{cases} \quad (1)$$

where $\lambda_N$ is the spin diffusion length in N, $e$ is the electron charge, $\lambda_{EE}$ is the Edelstein length and $\boldsymbol{E}$ is the external electric field. The spin current in region N1 and N2 reads:

$$\begin{cases} \boldsymbol{j}_s^{N1}(z) = -\frac{1}{2eR_N A}\left[-\boldsymbol{a}e^{-\frac{z-d_2}{\lambda_N}} + \boldsymbol{b}e^{\frac{z-d_2}{\lambda_N}}\right] - \theta_1 j_{c1}\boldsymbol{e}_y \\ \boldsymbol{j}_s^{N2}(z) = -\frac{1}{2eR_N A}\left[-\boldsymbol{c}e^{-\frac{z}{\lambda_N}} + \boldsymbol{d}e^{\frac{z}{\lambda_N}}\right] - \theta_2 j_{c2}\boldsymbol{e}_y \end{cases} \quad (2)$$

where $R_N = \rho_N\lambda_N/A$, and $\lambda_{EE} = \alpha_R\tau/\hbar$. $\lambda_{EE}$, $\alpha_R$ and $\tau$ are Rashba parameter and effective momentum scattering time. The spin current going through F/N2 interface reads:

$$\boldsymbol{j}_s' = \frac{G_{\uparrow\downarrow}'}{2eA_J}\left[\widehat{\boldsymbol{M}} \times \left(\widehat{\boldsymbol{M}} \times \delta\boldsymbol{\mu}(0)\right)\right] + \frac{G_{\uparrow\downarrow}''}{2eA_J}\left(\widehat{\boldsymbol{M}} \times \delta\boldsymbol{\mu}(0)\right) \quad (3)$$

where the spin mixing conductance $G_{\uparrow\downarrow} = G_{\uparrow\downarrow}' + iG_{\uparrow\downarrow}''$. Solving Eq. (1), (2) and (3), we obtain the spin current and spin accumulation in each layer, which are used to calculate an additional longitudinal charge current induced by the inverse SHE and REE. In the inverse REE, $\delta j_c^{N2} = \lambda_{EE}(\sigma_N/4e\lambda_N^2)\delta\mu_{N2}^y$ is assumed. The overall MR due to the direct and inverse SHE and REE:

$$\frac{\Delta R_{xx}}{R_0} \approx -\frac{\lambda_N/\rho_N}{d_N/\rho_N + d_F/\rho_F}\left(\theta_{SH}\Lambda + \frac{\lambda_{IEE}}{2\lambda_N}\Phi\right)^2 \mathrm{Re}\left[\frac{g_s}{1+\Gamma g_s}\right]$$

where

$$\begin{cases} \Lambda = \tanh\left(\frac{d_\mathrm{N}}{2\lambda_\mathrm{N}}\right) \\ \Phi = \begin{cases} \left[1 - \frac{\sinh\left(\frac{d_\mathrm{N} - d_2}{\lambda_\mathrm{N}}\right)}{\sinh\left(\frac{d_\mathrm{N}}{\lambda_\mathrm{N}}\right)}\right] \text{ for } d_\mathrm{N} > d_2 \\ 1 \text{ for } 0 < d_\mathrm{N} < d_2 \end{cases} \\ \Gamma = \coth\left(\frac{d_\mathrm{N}}{\lambda_\mathrm{N}}\right) \\ g_\mathrm{s} = 2\rho_\mathrm{N}\lambda_\mathrm{N}G_{\uparrow\downarrow} \end{cases}$$

Considering an exponential decay in the REE spin accumulation, the overall MR reads:

$$\begin{aligned} \frac{\Delta R_{xx}}{R_0} \approx & -\frac{\lambda_\mathrm{N}/\rho_\mathrm{N}}{d_\mathrm{N}/\rho_\mathrm{N} + d_\mathrm{F}/\rho_\mathrm{F}} \mathrm{Re}\left[\frac{2\rho_\mathrm{N}\lambda_\mathrm{N}G_{\uparrow\downarrow}}{1 + 2\rho_\mathrm{N}\lambda_\mathrm{N}G_{\uparrow\downarrow}\coth(d_\mathrm{N}/\lambda_\mathrm{N})}\right] \\ & \times \left\{\left[\theta_\mathrm{SH}\tanh\frac{d_\mathrm{N}}{2\lambda_\mathrm{N}} + \left(\frac{\lambda_\mathrm{IEE}}{2\lambda_\mathrm{N}}\right)\right]^2 \right. \\ & - 2\left(\frac{\lambda_\mathrm{IEE}}{2\lambda_\mathrm{N}}\right)\left[\theta_\mathrm{SH}\tanh\frac{d_\mathrm{N}}{2\lambda_\mathrm{N}} + \left(\frac{\lambda_\mathrm{IEE}}{2\lambda_\mathrm{N}}\right)\right]\frac{\langle\sinh[(d_\mathrm{N} - d_2)/\lambda_\mathrm{N}]\rangle}{\sinh(d_\mathrm{N}/\lambda_\mathrm{N})} \\ & \left. + \left(\frac{\lambda_\mathrm{IEE}}{2\lambda_\mathrm{N}}\right)^2 \frac{\langle\sinh^2[(d_\mathrm{N} - d_2)/\lambda_\mathrm{N}]\rangle}{\sinh^2(d_\mathrm{N}/\lambda_\mathrm{N})}\right\} \end{aligned}$$

where

$$\langle\sinh[(d_\mathrm{N} - d_2)/\lambda_\mathrm{N}]\rangle = \frac{\sinh(d_\mathrm{N}/\lambda_\mathrm{N}) - r\cosh(d_\mathrm{N}/\lambda_\mathrm{N}) + re^{-(d_\mathrm{N}/\lambda_\mathrm{N})/r}}{(1 - r^2)[1 - e^{-(d_\mathrm{N}/\lambda_\mathrm{N})/r}]}$$

$$\langle\sinh^2[(d_\mathrm{N} - d_2)/\lambda_\mathrm{N}]\rangle = \frac{\cosh(2d_\mathrm{N}/\lambda_\mathrm{N}) - 2r\sinh(2d_\mathrm{N}/\lambda_\mathrm{N}) - e^{-(d_\mathrm{N}/\lambda_\mathrm{N})/r}}{2[1 - (2r)^2][1 - e^{-(d_\mathrm{N}/\lambda_\mathrm{N})/r}]} - \frac{1}{2}$$

where $\langle\cdots\rangle$ denotes the average with respect to $d_2$ weighted by $\exp(-d_2/d_\mathrm{R})$, $r \equiv d_\mathrm{R}/\lambda_\mathrm{N}$, where $d_\mathrm{R}$ is the effective REE thickness of an equivalent layer where the spin accumulation density is homogenous.

## Supplementary Note 3. On the magnitude of $\lambda_{\mathrm{IEE0}}$ and Rashba-Edelstein magnetoresistance

In the following, we briefly discuss the $\lambda_{\mathrm{IEE0}}$ values for our epitaxial Pt/Co bilayer interface together with the REE component of the MR signal (*i.e.* the REMR). We start with one of the most prominent Rashba interface: the Bi/Ag interface [4,5]. A previous study on Bi/Ag/CoFeB trilayers shows a maximum REMR of about 0.05% with a $\lambda_{\mathrm{IEE0}}$ of 0.3 nm [6]. Such a $\lambda_{\mathrm{IEE0}}$ value is comparable to that obtained in Bi/Ag/NiFe trilayers via spin pumping [4]. As suggested by Eq. (2) in the main text, the magnitude of the REMR scales with the squares of $\lambda_{\mathrm{IEE0}}$. By assuming that both Bi/Ag and Pt/Co interfaces have the same REMR–$\lambda_{\mathrm{IEE0}}$ scaling, one expects a large REMR value of 1.3–6.8% in epitaxial Pt/Co bilayers. This is in fair agreement with the MR values of 0.5–2.3% as observed in this study shown in Fig. 2. We note here that this is a very rough estimation because the assumption of the same REMR–$\lambda_{\mathrm{IEE0}}$ scaling for the two Rashba interfaces requires, *e.g.* that the two interfaces have the same spin relaxation length $\lambda_{\mathrm{sf}}$ and Rashba thickness $d_2$, which has not been verified yet. Such a discussion only shows that the extracted $\lambda_{\mathrm{IEE0}}$ of 1.5–3.5 nm are plausible values for the our epitaxial Pt/Co interface.

## Supplementary Note 4. Thickness calibration for wedged thin films at 300 K

The Pt thickness of the wedged thin films is calibrated by comparing their room-temperature inverse sheet resistance $[1/R_{xx}^{0} * (l/w)]$ with those of flat thin films. Here we briefly explain the determination of the Pt thickness of wedged samples: wedged Pt thin films were prepared by using a linear shutter (inside the sputtering chamber) that was moving at a constant speed of 0.1 mm/s. Assuming the sputtering rate of Pt to be A nm/s, we obtain the gradient of the wedged thin films to be 10*A nm per mm. Note that the neighboring Hall bars along wedge direction on one 1 cm*1 cm substrate have a fixed distance (let it be B mm), the neighboring Hall bar devices thus have a Pt thickness difference of 10*A*B nm. Because the sheet resistance of Hall bar devices along the wedge direction varies monotonously, we are able to calibrate the thickness of wedged Hall bar devices using a flat film Hall bar device. Figure S3 shows the inverse sheet resistance of epitaxial Pt/Co bilayers plotted as a function of Pt thickness where Co thickness ranges from 0.6 nm to 10 nm. The black squares denote the results of wedged samples and the circles denote those of flat thin films.